\documentstyle[12pt]{article}
\begin{document}
\newcommand{\ot}{\frac{1}{2}}
\newcommand{\D}{& \displaystyle}  
\newcommand{\di}{\displaystyle}   
\font \math=msym10 scaled \magstep 1
\newcommand{\vecu}{{\bf u}}
\newcommand{\vecv}{{\bf v}}
\newcommand{\ZETA}{\mbox{\boldmath $\zeta$}}
\newcommand{\LL}{\ell}
\newcommand{\kappacrit}{\kappa_{\mbox{\scriptsize crit}}}
\newcommand{\geff}{g_{\mbox{\scriptsize eff}}}
\newcommand{\dising}{\delta_{\mbox{\scriptsize Ising}}}
\newcommand{\dres}{\delta_{\mbox{\scriptsize Res}}}
\newcommand{\dtotal}{\delta_{\mbox{\scriptsize total}}}

\date{May 1992}
\title
{Resonance Scattering Phase Shifts in a 2-d Lattice Model
\thanks{Supported by Fonds zur F\"orderung der
Wissenschaftlichen Forschung in \"Osterreich, project P7849.} }
\author
{\bf C.R. Gattringer\thanks{Present address:
Max Planck Institut f\"ur Physik und Astrophysik,
F\"ohringer Ring 6, D-8000 Munich, Germany}
and C.B. Lang \\  \\
Institut f\"ur Theoretische Physik,\\
Universit\"at Graz, A-8010 Graz, AUSTRIA}
\maketitle
\begin{abstract}
We study a simple 2-d model representing two fields with different mass
and a 3-point coupling term. The phase shift in the resonating
2-particle channel is determined from the energy spectrum obtained in
Monte Carlo simulations on finite lattices. Masses and wave function
renormalization constants of the fields as well as mass and width of the
resonance are determined and discussed. The representation of scattering
states in terms of the considered operators is analysed.
\end{abstract}

\newpage
\section{Introduction and motivation}

Determination of scattering phase shifts is still considered one of the open
problems in the nonperturbative lattice approach to quantum field theory.
Up to now bulk quantities, masses and matrix elements were the central
objects. Scattering processes involve at least four asymptotic states;
in QCD each of them is a bound state in terms of the original fields,
indeed a very demanding situation. However, various contributions to
this question exist \cite{LuCMP104}-\cite{GaLa91}.

The problem is particularly pressing, as most of the hadronic states to be
analysed are resonances and hence not asymptotic states of the field
theory.
This makes any rigorous control of finite size effects, which
necessarily turn up due to numerical limitations, extremely difficult.
Furthermore, phase shifts provide the only direct comparison to experiment.
Resonance parameters are inferred from phase shifts whereas positions
of resonance
poles in the unphysical sheet are plagued by necessary assumptions
implicit in the phenomenological analysis.

In a series of papers L\"uscher \cite{LuCMP104}-\cite{Lu91a}
introduced a method to obtain scattering
phase shifts from an analysis of the energy spectrum in a finite size
system. Eventually this method should be used in the full 4-d
theory
\cite{Lu91b}, but before we can hope to attack the determination of
phase shifts in lattice QCD it is appropriate to deal with simpler
models. L\"uscher and Wolff \cite{LuWo90} studied the O(3)
non-linear sigma
model in d=2, where analytic results for the scattering phases exist.
That model may be interpreted as describing scattering of three
mass-degenerate particles with certain symmetries. In order to discuss a
situation of two fields with different masses and potentially
resonating scattering behaviour we attacked another simple 2-d model
\cite{GaLa91}. Here we present a full account of our analysis and
results.

The organisation of this paper is as follows. In the next section
we briefly review L\"uscher's approach for the calculation of scattering
phases in a two dimensional lattice field theory. We also
discuss the idea for the  numerical determination of the scattering
spectrum from Monte Carlo data as presented in \cite{LuWo90}. Based on
this, we formulate
a method that allows one to compute the contribution of a lattice
operator to physical states. In the third section we present our model
and discuss
basic properties and our Monte Carlo simulation. The fourth
section contains results for the scattering parameters.
A determination of the wavefunction renormalization constants provides
better understanding of the observed
dependence of the resonance width on the bare 3-point coupling.
This is followed by a section where we present our results for the
contributions of the lattice operators to the scattering states.
We end with some concluding remarks.

\section{Determination of scattering parameters}
\subsection{Energy spectrum and phase shift}

Let us start with a brief review of L\"uscher's approach
\cite{LuCMP104}-\cite{LuWo90}. Throughout the paper we consider
the dimensional quantities as given in units of the appropriate
powers of the lattice spacing $a$. Consider the scattering
of two bosons of mass $m$ in a two dimensional system of finite spatial
extension $L$, but infinite time extension.
In this finite volume the momentum of a particle can only have
discrete values. Assume that the interaction region is localized and
smaller than $L$.
Imposing periodic boundary conditions, the quantization condition for
the relative momentum of two scattering particles reads
\begin{equation}\label{Phaseshiftrelation}
2 \delta(k_n) + k_n L =  2 n \pi , \quad n \in {\mbox{\math N}} ,
\end{equation}
where $\delta(k)$ denotes the phase shift acquired in the interaction
region. If its functional form is known, one could use
this relation to find the quantized values $k_n$. On the
other hand, given the momentum spectrum, (\ref{Phaseshiftrelation})
allows the determination of $\delta(k)$ for each $k$. This approach may
be realized by utilizing the dispersion relation for the energy of two
particles (of equal mass $m$) with relative momentum $k$
\begin{equation} \label{Wdispersionrelation}
W = 2 \sqrt{m^2+k^2} \quad .
\end{equation}
The energy spectrum in the two particle sector may be computed in a
Monte Carlo simulation for a lattice with spatial extension $ L$.
Indeed, for given $L$ one measures a set of discrete values $W_n$,
thus $k_n$ and with (\ref{Phaseshiftrelation}) the corresponding
value $\delta(k_n)$. Repeating  this procedure for various lattice
sizes it is possible to find the phase shift for a large
set of momentum values.

The power of this approach lies in  the utilization of finite
size effects for the determination of physical quantities. One will have
to respect carefully the limitations. The interaction region
and the single particle correlation length ought to be smaller than the
spatial volume, in particular $1 / m << L$. Polarization effects due to
virtual particles running around the torus should be under control.
Lattice artifacts will turn up in $O(a^2)$ corrections, i.e.
for large values of $k$. For the determination of the energy spectrum
one should consider correlation functions of a sufficiently large number
of observables  with the correct quantum numbers, capable to represent
the space of scattering states \cite{LuWo90}.

\subsection{Correlation functions}

We compute the connected cross-correlations
\begin{equation} \label{Crosscorrelation}
M_{n m} ( t ) :=
\langle {N_n}^*(0) N_m (t) \rangle_c =
\langle {N_n}^*(0) N_m (t) \rangle -
\langle {N_n}^* \rangle \langle N_m \rangle
\end{equation}
of operators in the scattering sector. Each $N_n, n = 1,2...$ is defined
on a single time slice, and $t$ denotes the separation of the time
slices. Ideally we have to consider a  complete set of states. The
explicit form of the operators will be given in section 4, after
introduction of the model.

The transfer matrix formalism (see e.g. \cite{ItDr89}) yields
the spectral decomposition
\begin{equation} \label{Spectraldeco}
M_{n m} ( t ) = \sum_{l=1}^{\infty} {v^{(l)}_n}^* v^{(l)}_m e^{- t W_l}
\quad .
\end{equation}
For simplicity we assume non-degenerate $W_l$, ordered
increasingly. The amplitudes $v^{(l)}_n = \langle l | N_n 0 \rangle$
are the projections of the states $| N_n 0 \rangle$ (generated by the
operators $N_n$ out of the vacuum) on the energy eigenstates
$\langle l|$ of the scattering problem.

For the determination of the energy spectrum from
(\ref{Spectraldeco}) we follow \cite{LuWo90}.
On a finite lattice, in the elastic regime below the
four particle threshold (or the three particle threshold,
depending on the model) there is a finite number of energy
eigenstates. Only two-particle states contribute and the
projections onto physical single particle states vanish.
Also, in a Monte Carlo calculation one
can only take into account a finite number $r$ of operators, which,
however should exceed the considered number of eigenstates.

One can expect that the spectral decomposition of $M(t)$ is approximated
rather well by the truncated matrix
\begin{equation} \label{Spectraldecotrunc}
C_{n m} ( t ) = \sum_{l=1}^{r} {v^{(l)}_n}^* v^{(l)}_m
e^{-t W_l} \;\; ,
\quad n,m = 1,2\ldots r  \quad .
\end{equation}
The generalized eigenvalue problem
\begin{equation} \label{Geneigprob}
C( t ) \; {\ZETA}^{(k)}(t) = \lambda^{(k)} ( t , t_0 ) \;
C( t_0 ) \;
{\ZETA}^{(k)}(t) \;\; ,
\hspace{0.5cm} k = 1,2, .. , r \hspace{0.5cm}.
\end{equation}
allows one to find the energy eigenvalues $W_l$ efficiently even for
not too large values of $t$.
In (\ref{Geneigprob}) we assume $t_0<t$, e.g. $t_0=1$.
We require the $N_n \; , \; n = 1,2,..,r$ to be linearly independent,
thus $C( t_0 )$ is regular and the generalized eigenvalue problem is
well defined. Indeed, it may be transformed to a standard eigenvalue
problem for the matrix
\begin{equation} \label{matrixT}
C^{-\ot} ( t_0 ) \; C ( t ) \; C^{-\ot} ( t_0 ),
\end{equation}
with identical eigenvalues $\lambda^{(k)} ( t , t_0 )$
and eigenvectors
${\vecu}^{(k)} (t) = C^{\ot} ( t_0 ) {\ZETA}^{(k)}(t)$.

The solution for the eigenvalues is
\begin{equation} \label{Eigenvalue}
\lambda^{(k)} ( t , t_0 ) = e^{- ( t - t_0 ) W_k} \quad .
\end{equation}
This can be easily verified by inserting it into (\ref{Geneigprob}),
leading to
\begin{equation} \label{Zetaequ}
B^{(k)}(t) {\ZETA}^{(k)} (t) = 0 \quad ,
\end{equation}
where
\begin{equation} \label{Bmatrix}
B^{(k)}_{nm} ( t ) = \sum_{l \neq k } {v^{(l)}_n}^* \; \Big[
v^{(l)}_m \Big( e^{- t W_l} - e^{- t_0 W_l} e^{- ( t - t_0 ) W_k}
\Big) \Big] \quad .
\end{equation}

Each row of the matrix $B(t)$ is a
linear combination of the $r-1$ linearly
independent vectors ${\vecv}^{(l)} , l = 1,2,..,k-1,k+1,..,r \; .$
Thus the space of the solutions of
(\ref{Zetaequ}) has dimension 1, as is needed to construct a normalized
eigenvector for the corresponding standard eigenvalue problem.
The effect of the truncation was shown to behave like
$\exp [ -(t-t_0) W_{r+1} ]$ by a
perturbation calculation \cite{LuWo90}.

To summarize, the determination of the spectrum in the
scattering sector consists of the following steps. Compute the
correlation matrix $M ( t )$ for various values of $t$
and calculate
$M^{-\ot} ( t_0 ) M ( t ) M^{-\ot} ( t_0 )$.
The dimension $r$ of $M$ cannot  be chosen too large, because then the
inversion turns out to be numerically unstable. Finally solve the
eigenvalue problem for this matrix, and determine the spectrum
from the exponential decay (\ref{Eigenvalue}).

\subsection{Eigenstates and spectral density}

Like in any Quantum Mechanics application the central problem is to
consider a set of operators with a sufficient overlap with the energy
eigenstates. We now demonstrate that the eigenvectors
${\ZETA}^{(k)}$ of the
generalized eigenvalue problem contain information about the
representation of the physical states by the chosen operators $N_n$.

Rewriting (\ref{Zetaequ}) gives
\begin{equation}
\sum_{l \neq k } {\vecv}^{(l) \dag } \Big[
\Big( e^{- t W_l} - e^{- t_0 W_l} e^{- ( t - t_0 ) W_k}
\Big) {\vecv}^{(l)} \cdot {\ZETA}^{(k)}(t) \Big] = 0
\end{equation}
for all $k$. Since the ${\vecv}^{(l) \dag}$ are linearly independent,
their coefficients have to vanish, yielding
${\vecv}^{(l)} \cdot {\ZETA}^{(k)} (t)  = 0 \; \forall \; l \neq k \; .$
Thus ${\ZETA}^{(k)} (t) \propto  {\vecv}^{(k)}$ and, when  normalized
to unit length, independent of $t$. Introducing the normalized eigenvector
${\ZETA}^{(k)} = {\vecv}^{(k)} / \mid \vecv^{(k)} \mid $ also the
eigenvectors ${\vecu}^{(k)} = C^{\ot} ( t_0 ) {\ZETA}^{(k)}$
of the associated standard eigenvalue problem are time independent.

We may express the original approximation (\ref{Spectraldecotrunc})
to $M(t)$ in terms of the obtained eigenvectors,
\begin{equation} \label{NewSpectraldecotrunc1}
C_{n m} ( t ) = \sum_{l=1}^{r} {\zeta^{(l)}_n}^* \zeta^{(l)}_m
\mid \vecv^{(l)}\mid ^2 e^{-t W_l} \;\; , \quad n,m = 1,2\ldots r
\quad .
\end{equation}
In particular consider the diagonal element
\begin{equation} \label{NewSpectraldecotrunc2}
C_{nn} ( t ) = \sum_{l=1}^{r} \mid \zeta^{(l)}_n\mid ^2
  \mid \vecv^{(l)}\mid^2 e^{-t W_l} \quad .
\end{equation}
For a continuous energy spectrum,
\begin{equation} \label{ContSpectraldecotrunc}
C_{nn}(t) = \int dW \rho_n(W) e^{-t W} \quad,
\end{equation}
and we identify  $\mid \zeta^{(l)}_n\mid ^2 \mid \vecv^{(l)}\mid^2
\propto \rho_n(W_l)$, the spectral density of the
correlation function of operator $N_n$.
We therefore find that $\mid \zeta^{(l)}_n\mid ^2$ indicates the
relative weight of the contribution of operator $N_n$ to the energy
eigenstate $\mid l  \rangle$.

In the calculation again one has to deal with the correlation matrix
$M$ and not with the truncated form $C$. This implies that results
obtained from ${\ZETA}^{(l)} := M^{-\ot} ( t_0 ) \; {\vecu}^{(l)}$,
where ${\vecu}^{(l)}$ are the eigenvectors of
$ M^{-\ot} ( t_0 ) M ( t ) M^{-\ot} ( t_0 )$, are affected by a term
$O \Big( e^{- t W_{r+1}} \Big)$, where $r$ is the rank of $M$.

If energy eigenvalues are degenerate, the procedure does not apply
and the amplitudes cannot be identified in a unique way. The method
then gives equal importance for the involved operators.

\section{Model and Monte Carlo simulation}
\subsection{Model}
We choose a model, which describes two light particles $\varphi$ that
couple to a heavier particle $\eta$ giving rise to resonating behaviour.
For this we couple two Ising fields through a 3-point term. The
action is given by
\begin{eqnarray}\label{action}
S &=\D -\kappa_\varphi \sum_{x \in \Lambda, \mu=1,2} \varphi_x
\varphi_{x+\hat{\mu}} \nonumber \\
  & \D -\kappa_\eta \sum_{x \in \Lambda, \mu=1,2} \eta_x \eta_{x+\hat{\mu}}
\nonumber \\
  & \D + \frac{g}{2}\sum_{x \in \Lambda, \mu=1,2} \eta_x \varphi_x
            (\varphi_{x-\hat{\mu}} + \varphi_{x+\hat{\mu}})   \quad .
\end{eqnarray}
Here, $\hat{\mu}$ denotes the unit vector in direction $\mu$.
The values of the fields are restricted to $\{ +1, -1\}$. Quantization
corresponds to taking expectation values with the measure
$Z^{-1} \exp{(-S)}\: d\varphi d\eta$ where $Z$ denotes the normalizing
partition function. The sum runs
over all sites $(x_0, x_1)$ of the euclidean $L\times T$ lattice
$\Lambda \subset {\mbox{\math Z}}_L \times {\mbox{\math Z}}_T$
and we imply periodic boundary conditions.
The 3-point term was introduced in a nonlocal but symmetric way,
because $ \varphi_x^2 \equiv 1$.

The case $g = 0$ corresponds to the situation of two independent Ising
models, each with a 2nd order phase transition at
$\kappacrit =\ot \ln (1+\sqrt{2}) \simeq 0.44068$.
In the scaling limit the model describes an interacting boson with mass
$m = - \log ( \tanh \kappa) - 2 \kappa$ \cite{ItDr89}.
Reformulating  the Ising model at the
phase transition as a theory of non-interacting
fermions, it was shown that the scattering matrix assumes the value $- 1$
independent of the momentum \cite{SaMiJi77}.

In the coupled case ($g > 0$), if we identify $\eta$ and $\varphi$ with
particle states, the term proportional to $g$ gives rise to
transitions like $\eta \rightarrow \varphi \varphi$ rendering $\eta$ a
resonance in the $\varphi$ channel, when kinematically allowed.
The notion of mass is then no longer well defined for the $\eta$ as it is no
longer an asymptotic state of the theory. For us, this is just the
situation of interest. Henceforth, whenever we say ``mass'' of the
$\eta$ field we really mean the position where the phase shift assumes
its resonating value. For given $g$, ``masses'' of the fields $\eta$ and
$\varphi$ may be adjusted by calibrating the hopping parameters
$\kappa_\varphi$ and $\kappa_\eta$.

\subsection{Phase structure}
In the following we give some simple arguments about
the phase structure of our model in the ($\kappa_\varphi$,
$\kappa_\eta$, $g$) space (in the thermodynamic limit).

For $g = 0$ we have the Ising phase transitions for $\kappa_\varphi =
\kappacrit$ and  $\kappa_\eta=\kappacrit$. In the general case the model is
even in the $\varphi$ field. In the symmetric
phase of the $\eta$-field there is a symmetry $g \leftrightarrow -g$.

For $g > 0$ we first consider the situation with a quenched $\eta$
configuration and discuss {\em the dynamics of} $\varphi$. The action
then has just a hopping term for
the $\varphi$-field with a locally varying coupling
\begin{equation}\label{kappalphilocal}
\omega_{x,\mu} =
\kappa_\varphi - \ot g (\eta_x + \eta_{x+\hat{\mu}}) \quad .
\end{equation}
For sufficiently small $g$ this is still a ferromagnetic system and the
Griffiths inequalities apply. Thus all link expectation values are
monotonically increasing with $\omega_{x,\mu}$. For sufficiently large
$\kappa_\varphi$ there will be an ordered and for sufficiently
small $\kappa_\varphi$ a disordered phase. This
situation (Potts model with quenched random bonds) has been recently
studied in detail \cite{Ai89,HuBe89} where the authors prove that
there is a second order phase transition.
To our knowledge it is unclear, whether this behaviour survives
for a dynamical $\eta$ field.

In order to examine {\em the $\eta$ dynamics} let us discuss
two limiting cases. For $\kappa_\varphi = \infty$ the $\varphi$ spins
are strictly ordered. Then the three-point term in
(\ref{action}) effectively acts
like an external magnetic field of strength $g$. For non-zero $g$ the
field $\eta$ has long range order and does not undergo a phase
transition.
On the other hand for $\kappa_\varphi=0$, $\varphi$ is completely
disordered and the interaction term acts like an external random field
for the $\eta$-spins.
Random magnetic fields are expected to lower the effective
dimensionality (i.e. raise the critical dimension by 2)
\cite{ImMa75}-\cite{DaAn84}.
Recently it was proved \cite{Ai89} that for a quenched random magnetic
field  the field  $\eta$ remains disordered for all values of the
hopping parameter. Hence we conclude that there is no phase transition
in the $\eta$ field for finite $\kappa_\eta$ and  $\kappa_\varphi=0$,
$g>0$.

We expect therefore that the $\eta$ phase transition extending at $g=0$
over all values of $\kappa_\varphi$ disappears at $g>0$, at least for
the edges $\kappa_\varphi= 0, \infty$. Our numerical results for
masses and bulk quantities show that the phase transition
becomes weaker for $g>0$, changing into a cross-over like behaviour
with large, but finite correlation length at the $\eta$-transition.

To summarize these arguments, one cannot be sure that for $g\neq 0$ the
model (\ref{action}) undergoes a second order phase transition for both
fields. This would indicate that the cutoff cannot be removed completely
for $g>0$ and that there is no continuum limit possible, in which both
fields describe mutually interacting physical particles with finite
mass. We are always dealing with an effective model.

However, we are not presenting a model for a possible continuum field
theory. All we want is to study the possibility of extracting phase
shifts in a situation where two particles of different mass have an
effective interaction.
For this aim the lattice is considered more a technical means to
approximate an underlying continuum model than a method for a
rigorous construction of a continuum limit.
In the range of bare couplings considered the model serves our purpose.
However, not all combinations of correlation lengths (i.e. mass values)
can be realized. In particular a study of lattice artifacts is not
possible for arbitrary small lattice spacing.

\subsection{Cluster algorithm}

For the Monte Carlo simulation of Ising models the cluster algorithm
\cite{SwWa87} is a very powerful tool, both for generating new
configurations and for defining and measuring improved estimators
\cite{Sw83}. In our case with a 3-point term we have to modify it.
We update the $\eta$ and the $\varphi$ field alternately.

{\bf Updating the $\eta$ spins:~} Bonds between neighboured
spins of equal sign are kept with the probability
$1-\exp{(-2\kappa_\eta)}$. After identification of the connected
clusters all clusters are flipped, each with separately determined
probability
\begin{eqnarray}\label{flipetacluster}
p^{flip}_\eta &=\D 1/\left(1+ e^{-2\alpha (C)}\right), \nonumber \\
\alpha (C) &=\D \frac{g}{2} \sum_{x\in C,\mu=1,2}  \eta_x \varphi_x
            (\varphi_{x-\hat{\mu}} + \varphi_{x+\hat{\mu}}) .
\end{eqnarray}
This method has been suggested \cite{SwWa87} in order to incorporate
the effect of constant external fields.

{\bf Updating the $\varphi$ spins:~} This part of the action is treated
like an Ising system with the locally varying hopping parameter
$\omega_{x,\mu}$ introduced in (\ref{kappalphilocal}).
In identifying the clusters we keep  bonds between like-sign neighbours
alive with probability $1 - \exp{(-2\omega_{x,\mu})}$ and flip clusters
with probability $\ot$.

The cluster method is ergodic and fulfills the detailed balance
condition; for the values of $g$ considered it is a clear improvement
over the Metropolis updating.

We study the model at three values of the 3-point coupling
$g$ = 0, 0.02 and 0.04; for later reference we denote these cases by I, II
and III. Our first step was to tune the hopping parameters.
We measured single particle propagators to
adjust the $\varphi$-mass and find a crude estimate for the
energy of the $\eta$ resonance. We tried to establish the values
$m_\varphi \simeq 0.19$ and $m_\eta \simeq 0.5$.
Table 1 gives the couplings where we performed systematic simulations
on various lattice sizes.

Throughout this work we use $T=100$; the spatial extension $L$ varies
between $12$ and $60$. For each set of couplings and lattice size we
performed typically $2\times 10^5$ measurements.
For the statistical errors
we display one standard deviation, estimated with the Jackknife
method (For a discussion see e.g. \cite{YaRo86}).

\section{Results for the scattering parameters}

For the determination of the energy spectrum a precise knowledge of the
single particle mass and related finite size effects is important.
In this section we first discuss our results for the single particle
state, then for the two particle channel, in particular the phase
shifts. Finally we perform a computation of
wavefunction renormalization constants, to clarify the scaling of the
resonance width with $g$.

\subsection{Single particle states}

Let us first study the single particle propagator. The operator of
a $\varphi$ state with momentum
\begin{equation}
p_{1,\nu} = 2 \pi \nu /L,\quad \nu = -L/2+1, \ldots, L/2
\end{equation}
may be represented through
\begin{equation}\label{Operatorphi}
\frac{1}{L}\sum_{x_1\in \Lambda_{x_0}}
\varphi_{x_0,x_1} \exp{(i x_1 p_{1,\nu})} \quad ,
\end{equation}
where $\Lambda_{x_0}$ denotes a timeslice of $\Lambda$. Its
connected correlation function over temporal distance $t$ decays
exponentially $\propto \exp{(-p_{0,\nu} t)}$ defining the
single particle energy $p_{0,\nu}$; in particular we have $p_{0,\nu=0} =
m_\varphi$.

The observed mass, as compared to the ``real'' mass at vanishing
lattice spacing and infinite volume, incorporates contributions from
polarization due to self interaction around the torus
(decreasing exponentially with L)
and lattice artifacts due to the finite ultra-violet cutoff
(polynomial in the lattice constant a).

In fig. 1 we plot $p_{0,\nu}$, obtained for the pure
Ising model (I) and $L=50$. We compare the values with
the continuum spectral relation ,
\begin{equation}
p_{0,\nu} = \sqrt{m^2 + p_{1,\nu}^2}
\end{equation}
and find deviations $O( (a p)^2)$, as expected \cite{LuWo90}.
We also plot the dispersion relation for
the lattice propagator of a Gaussian particle with mass $m$,
\begin{equation}
p_{0,\nu}= \mbox{arcosh}( 1-  \cos p_{1,\nu}+ \cosh m )\quad ,
\end{equation}
and find excellent agreement with the measured values. Thus
this particular single particle state shows little deviation from the
free lattice form. This results holds for models (II) and (III)
too.

The Ising model may be considered the limit of
a $\phi^4$-theory with infinite bare 4-point coupling $\lambda$.
We are in the symmetric phase. Thus we expect the leading interaction
to be due to an effective 4-point term.
The finiteness of the spatial volume allows self-interaction
with one particle running around the torus. This leads to
a finite volume correction to the particle mass.
Performing a perturbation calculation for two dimensional
$\varphi^4$ theory along the lines of \cite{Lu89} one finds a decrease
\begin{equation}\label{mdecrease}
M ( L ) - m \;\;\propto \;\;\lambda L^{-\ot} e^{- m L} +
O \Big(\lambda^2, L^{-\ot} e^{-2 m L} \Big)
\hspace{1.cm} ,
\end{equation}
where $M(L)$ is the mass measured on a lattice of spatial size $L$, and
$m$ is the infinite volume mass. A 3-point interaction leads to a different
exponential.

Fig. 2, again for model (I), gives our results for
$ \ln [ \sqrt{L} (M(L) - m)]$ vs. $L$
and indicates
perfect agreement with (\ref{mdecrease}). Indeed  the
fit to a linear behaviour gives a slope of $-1.047(31) m$ for case (I)
and $-1.015(25) m$ for (III), in good agreement with the expected
value $-m$.
For situation (I) this is obvious, because we are in the symmetric
phase where there is no 3-point coupling. In cases (II) and (III)
the $\eta$ particle that enters the three-point
interaction is much heavier than the $\varphi$ and its contribution is
therefore suppressed. This mass shift on finite lattices
has also been confirmed in Monte Carlo simulations of
the Ising model in 4 dimensions \cite{MoWe87}.

As already discussed in section 2, the infinite volume mass is explicitly
known in the Ising model. We compare
the measured value at $L=50$ for $g = 0$ with this exact result and
find a shift of
$+0.00006(13)$. We use this shift to estimate the infinite volume
mass values for the light particles in the coupled cases by subtracting
it from the values measured at $L=50$, correspondingly
(Anyway this shift lies within the errors).

Equivalencing the Ising system with an effective field theory of
renormalized fields, the real valued
variables $\Phi$ with the canonical continuum form of the kinetic and
the mass term for bosons, transform to the Ising spins $\phi$ via
$\Phi = \sqrt{\kappa} \phi$. The momentum space propagator for
the $\Phi$ fields is
\begin{equation}\label{Zdef}
G(p) = \frac{Z}{m_r^2 + p^2} \quad \mbox{for} \quad
p \longrightarrow 0 \quad ,
\end{equation}
where $m_r$ is the renormalized $\Phi$ mass and $Z$ the wave function
renormalization constant. It may be determined from the susceptibility
\begin{equation}\label{Z}
Z = \kappa  \; m^2 \frac{1}{L T}
\sum_{x1,x2 \in \Lambda} \langle \phi_{x_1} \phi_{x_2} \rangle_c
\quad .
\end{equation}
The renormalized field is given through
$\Phi_r = \Phi / \sqrt{Z} = \sqrt{\kappa} \; \phi / \sqrt{Z}$.

Since the definitions for $\eta$ and $\varphi$ are identical,
the symbol $\phi$ stands for either spin variable $\eta$ or
$\varphi$. We use the high temperature results \cite{SyGaRo72}
for the susceptibility in the Ising model, to check the correctness of
our numerical results for $Z$ in the uncoupled case.

Only  $Z_{\varphi}$ is well defined, since only
$\varphi$ is an asymptotic state of the full theory.
However, in the discussion of our results we
want to compare the obtained value for the resonance width with the
tree-level predictions of an effective Lagrangian model; for this we
need to consider properly renormalized effective fields. Only for
the sake of this qualitatively interesting comparison we determine an
``effective'' w.f. renormalization constant $Z_{\eta}$, aware of the
ill-definedness of that quantity in case (II) and (III).

\subsection{Two particle sector}
\subsubsection{Operators and phase shifts}
In the 2-particle channel we consider operators
with total zero momentum and quantum numbers of the $\eta$,
\begin{eqnarray}\label{Operators}
N_1(x_0) &=\D \frac{1}{L}\sum_{x_1\in \Lambda_{x_0}} \eta_{x_1,x_0}
\quad , \\ N_j(x_0) &=\D \frac{1}{L^2}\sum_{x_1,y_1\in \Lambda_{x_0}}
e^{ip_j(x_1-y_1)} \varphi_{x_1,x_0} \varphi_{y_1,x_0} \quad , \\
&\D\mbox{with~}p_j = \frac{2\pi (j-2)}{L} ,\quad j=2,3,\ldots
\nonumber
\end{eqnarray}
and measure all cross-correlations (\ref{Crosscorrelation}).
The operators $N_{j>1}$ describe two $\varphi$-particles in the
CM system with relative momentum $p_j$.
Due to the interaction they do not correspond to eigenstates of our
model. Indeed they are eigenstates of the Gaussian model of free bosons.
However, if the set is complete, the diagonalization of section 2
provides the necessary information on the energy spectrum of this
channel.

The number of operators considered
in (\ref{Crosscorrelation}) should be chosen larger than the number
of states in the elastic regime $2 m_\varphi \leq W < 4 m_\varphi$ and
not larger than $L/2$ to be linearly independent. A larger set provides
a better representation of the eigenstates but enhances the numerical
noise. We work with between 4 and 6 operators depending on $L$ (Only as
a testing case we once consider 8 operators for a particular run at
$L=60$, $g=0.02$). For the product
$M^{-\ot} ( t_0 ) M ( t ) M^{-\ot} ( t_0 )$.
we choose $t_0=1$;
for our statistics, usually the diagonalization becomes unstable around
$t = t_{\max}\simeq 5-8$, depending on $L$.
The values $W_\alpha$ are determined by averaging
$\log [\lambda_\alpha (t) /\lambda_\alpha (t+1)]$ for
$1 < t< t_{\max}$. For each lattice size we determine the
low lying part of the energy spectrum. Each value corresponds, via the
spectral condition (\ref{Wdispersionrelation}), to a value of $k$ and
with (\ref{Phaseshiftrelation}) to a phase shift $\delta(k)$.

For $g=0$ we have the situation of two independent Ising systems.
As discussed above, the critical Ising model in d=2 has an S-matrix
equal to $-1$ \cite{SaMiJi77}, and we therefore
expect a phase shift $\dising = -\frac{\pi}{2}
\pmod{\pi}$ in the scaling regime.
In fig. 3 we exhibit the energy spectrum (a) for a system of
non-interacting Gaussian fields (curves due to
(\ref{Phaseshiftrelation}) with phase shift $\delta = 0$)
and (b) as obtained in our simulation of the Ising model.
The curves for the two particle states in
fig. 3b are due to (\ref{Phaseshiftrelation}) for
$\delta = -\frac{\pi}{2}$ and we find excellent agreement. The
single particle state $\eta$ (horizontal line at $W=0.5$) is completely
decoupled from the $\varphi \varphi$ states and we therefore observe
degeneracies in the energy levels at certain values of $L$.
Fig. 4 gives the results for the phase shift $\delta (k)$ determined
from the data for the Ising model (fig. 3b) as a function of the
dimensionless momentum $k/m_\varphi$.

Already here we identify two sources for the relative magnitude of the
error bars. Small values of $k$ result from values of $W$ near
$2 m_\varphi$, at large $L$. Close to this threshold, due to the
functional dependence of (\ref{Wdispersionrelation}) the statistical
error in $W$ transforms into a relatively larger error of $k$ and thus
$\delta(k)$.
Higher $k$, on the other hand, stem from the large values of the
energy, where the statistical fluctuations are intrinsically
larger.

Figs. 5a and 5b exhibit results for $g\neq 0$. For most of the
points the errors are smaller than the symbols, typically $0.5\%$
of the energy value. The degeneracies in the energy spectrum have now
disappeared, resulting in avoided level crossing with a gap that grows
with $g$. The corresponding phase shift data are given in  fig.6.

One should keep in mind that points at neighboured values of
$k$ may come from quite different values of $L$ and different branches
of energy levels. In order to emphasize this situation we have
chosen corresponding symbols to denote energy values on one branch,
i.e. for one value of $n$. Each branch, from small to large
$L$ contributes to the whole range of  $k$ values; it starts at larger
values of $W$ (or $k$) with a phase shift close to $\pi/2$, passes
through the resonance value at the plateau and approaches $-\pi/2$
towards smaller $k$. The overall consistent behaviour is impressive.

The high energy values, corresponding to
large $k$ may feel the lattice cut-off effects $O(a^2)$ as well as
possible mis-representation of the energy eigenstates by the considered
operators. We discuss the second problem in a subsequent section. Within
the accuracy of our data none of these effects seems to be a problem.

\subsubsection{Resonance parameters}

The observed overall behaviour has a simple interpretation.
A resonance may be parametrized with the effective range approximation
\begin{equation}\label{EffRangeDelta}
\frac{k}{W} \cot{\dres} = a - b k^2 \quad  .
\end{equation}
At the resonance
$\dres \simeq \frac{\pi}{2}+ \frac{W-m_R}{\Gamma_R/2} $ ,
or
\begin{equation}
m_R = 2 \sqrt{(m_\varphi)^2 + (a/b)}, \quad
\Gamma_R =  \frac{4}{b\; m_R^2} \sqrt{\frac{a}{b}} \quad .
\end{equation}
Adding this resonance phase
to the Ising model background phase the spectrum
in the resonating channel should approximately  obey
\begin{eqnarray}\label{Delta}
\dtotal (k) &\equiv & \dising +
\dres(k) = -\arctan{\frac{a - b k^2}{k/W}} ,\\
\label{W}
W &=& 2 \sqrt{(m_\varphi)^2 + \bigl( -\frac{2}{L} \dtotal (k)
+ \frac{2 n \pi}{L}  \bigr)^2} \quad ,\quad n = 0,1,\ldots  \quad .
\end{eqnarray}
Here $n$ denotes the various energy levels in relation
(\ref{Phaseshiftrelation}). We superimpose this theoretically expected
behaviour of the spectrum  on the measured points in figs. 3-6.

The parameters $m_R\equiv m_\eta$ and $\Gamma_R\equiv \Gamma_\eta$ may be
determined from a direct fit to the shoulders in the energy curves. However,
this fit turns out to be not very stable and has much larger statistical
errors than a determination from the resulting phase shifts as discussed
below. We fit the values of $\tan \delta$ in the region
$-1.45 < \delta < 1.45$ to a straight line in $k^2$
according to (\ref{Delta}). This gives the effective range parameters
$a$, $b$ and our estimates for $m_\eta$, $\Gamma_\eta$ in table 1.
The resulting $\chi^2$ indicates that the statistical errors of our phase
shift data and the errors of $W$ may be underestimated (cf.
\cite{GaLa91}).

The ratio of the resonance widths for (III) and (II) is smaller than the
value $4$, the square of the ratio of the couplings $g$ that one naively
expects from the Born approximation. In doing that, however, we should
first write down an effective Lagrangian with properly renormalized
effective fields $\eta_r$ and $\varphi_r$, with kinetic parts and a
three-point term. For sake of such a comparison we have to
determine the wave  function renormalization constants $Z_\varphi$ and
$Z_\eta$, although the second is ill-defined as mentioned earlier
in sec. 4.1.
{}From this information we may estimate an effective three-point coupling
\begin{equation}
\geff = g \; \frac{1}{\kappa_\varphi \; \sqrt{ \kappa_\eta}} \;
Z_\varphi \sqrt {Z_\eta} \quad ,
\end{equation}
which may be a more suitable number to enter the tree-level formula.

With this naive procedure we obtain $Z_\eta$, $Z_\varphi$ and
$\geff$ for all three values of the bare coupling $g$ (see table 1).
Both $Z_\eta$ and $Z_\varphi$ are smaller than
1, even in the uncoupled  case. These values decrease for increasing
$g$. The values for $\geff$ are 0.0568(6) and 0.0938(8) for $g =
0.02$ and $g = 0.04$ respectively. With these values, the ratio
$\Big( \geff(0.04) \; / \; \geff(0.02) \Big)^2 =  2.73(11)$ is
closer to the observed ratio $\Gamma_\eta (0.04) \; / \; \Gamma_\eta (0.02)
= 2.13(26)$. The naive argument based on tree-level perturbation
theory appears to be at least qualitatively correct.  However, for a
better identification of our model with an effective field theory,
one has to take into account further interaction terms of higher order in
the fields. In fact the preferred definition of a renormalized
3-point coupling is just from the observed (physically well defined)
width.

\section{Representation of the physical states}

Is the number of lattice operators considered for the determination of the
energy levels in the scattering channel sufficiently large? To analyse the
representation of the physical states by the considered operators
(\ref{Operators}) we
compute the vectors ${\ZETA}^{(l)} = M^{-\ot} ( t_0 ) \;
{\vecu}^{(l)}$, normalized to unit length.
Then $\Big( \zeta^{(l)}_n \Big)^2$ provides a measure on how much $N_n$
contributes to the physical eigenstate $\mid l \rangle$. As discussed
in sect. 2.3 there should be no dependence on $t$. This is consistent
with the numerical results and therefore we average the values for all
available $t$.

Fig. 7 shows the relative weights $\Big( \zeta^{(l)}_n \Big)^2$ for
$l=1,2,3$ and $n=1,2,3$ for case (III). Due to the normalization of
${\ZETA}^{(l)}$ the individual contributions can assume values
between 0 and 1. If the three {\em displayed} operators are sufficient
to represent the considered state, their contributions should add up
to 1.

Since these values are displayed for all considered lattice sizes, a
remark concerning the number $r$ of components in ${\ZETA}^{(l)}$, which
is equal the rank of $M$, is appropriate. For small $L$, the lattice is
statistically ``nervous'', and the inversion of $M$ is possible only
for a small number $r$ of operators (e.g. $r=4$ at $L=12$, while $r=8$
at $L=60$). At small $L$, for the lowest energy eigenstate $| 1\rangle$
in the scattering sector mainly $N_1$ contributes (see the circles in
fig. 7a). Only $\zeta^{(1)}_1$ is
considerably different from zero. Hence for the lowest state $|1\rangle$
it makes no difference whether the vector ${\ZETA}^{(1)}$ is normalized
at length 3 or 4, because the 4th entry is very small anyway.
For the higher states $| l\rangle$ more than four entries of
${\ZETA}^{(l)}$ may be necessary. However, due to the numerical
instability of the inversion at small $L$, only four entries are
available.
Thus the results for the representation of the higher physical states
appear shifted for small $L$. For larger $L$ this technical
problem vanishes: The results for at least the three lowest eigenstates
are reliable.

Compare the energy spectrum of fig. 5b with fig.7.
Let us first discuss the lowest energy state $| 1\rangle$ ,
represented by circles in fig.5b, with respect to fig. 7a.
For small $L$, $W_1$ is close to the resonance energy. There the state is
completely dominated by the operator $N_1$ (circles in fig. 7a),
which describes an $\eta$-particle at rest.
For increasing $L$, $W_1$ decreases and approaches the 2-particle
threshold. This manifests itself in a drastic decline
of the contribution of $N_1$, accompanied by an increase of
the amplitudes for two $\varphi$-particles at rest (operator $N_2$,
triangles) and with relative unit-momentum (operator $N_3$, squares).
For lattice size larger than 20, the energy $W_1$ is already too far
below the resonance energy so that the energetically higher $\eta$ at
rest cannot contribute much to $| 1\rangle $. The state is built out
of $N_2$ and $N_3$ (with energies as shown in fig. 3a) almost
exclusively.

For the second lowest state $| 2\rangle $ (fig. 7b) the situation
is reversed. As seen in fig. 5b, for small $L$ the energy $W_2$ is
high above the resonance energy, so the $\eta$-state $N_1$ (circles)
does not contribute. The state is dominated by the lowest lying
2-$\varphi$ operators. For increasing $L$, $W_2$ decreases and as it
crosses the resonance energy, the contribution of $N_1$ peaks.
Leaving the resonance-plateau, state $| 2\rangle $ is again dominated by
the energetically favourable operator $N_3$. The contribution of $N_4$
is substantial as well, but not shown in the picture.

Fig. 7c gives the contributions to $| 3\rangle$.
In the considered $L$-region it does not show
any dramatic behaviour. For $L < 50$ the energy is too high for  $N_1$
and $N_2$ contributions and the state is built out of $N_3$ and higher
operators. Only for the largest $L$ the $\eta$-amplitude grows.
For small $L$ the remark concerning the rank of the correlation matrix
made earlier applies. Between $L=12$ and
$L=16$ there is a discontinuity in the $N_3$ contributions, due to
the discussed truncation effect. The diagonalization at $L=16$ includes
one operator more than that at $L=12$. This operator
has nonvanishing contributions and therefore affects the $N_3$ amplitude
via the normalization of ${\ZETA}^{(3)}$.

In the case (I) of two uncoupled Ising models we get a similar picture.
$N_1$ then measures the particle mass in the completely disconnected
$\eta$ system and does not affect the $\varphi$ states and vice versa.
The lowest eigenstates states of fig. 3b turn out to be mixtures
of the $\varphi \varphi$-states $N_n \; , \; n \geq 2$.

In particular, it is easy to identify misrepresentation of
energy eigenstates  in terms of the considered operators. At higher
energies one observes a shift of the weight  factors
$\Big( \zeta^{(l)}_n \Big)^2$ towards higher operators $N_n$.
Fig. 8 gives an example of a situation, where the center of weight has
moved to relatively high values of $n$. It is obvious, that more
operators should be included for a good representation of that
state.

Numerically we observe, that a truncation of the correlation matrix
$M(t)$ leads to a rise of the higher energy levels. However, in the
data presented here and used in the determination of the phase shift
this effect is under control. For the low lying energies it is
negligible and for the largest energy level it is still smaller than the
statistical error.

\section{Concluding remarks}

In our Monte Carlo study we have established, that L\"uscher's
suggestion for determining phase shifts is indeed a very
reliable method, at least in d=2. Our simple model describes a
system of two particles with a three-point coupling. The coupling
constants have been chosen such that one of the states may become a
resonance. Indeed we find resonating phase shifts in this channel.
As a byproduct we also obtain the phase shift of a pure Ising model.

The following remark concerns just the particular model studied here:
As discussed, for $g\neq 0$ we cannot expect to perform a continuum
limit. Most likely it is not possible to increase the correlation
length arbitrarily keeping the ratio $m_\eta /m_\varphi$ fixed.
Numerically we observed, that the $\eta$-contribution to the
peak in the specific heat becomes less pronounced for
increasing values of $g$ (This is the technical reason for not
studying even broader resonance widths). Therefore, in this model, it
might not be possible to study lattice artifacts in a more systematic
way.

The study of the representation of scattering states by the
operators entering the correlation functions sheds light on the
problem of constructing good approximations for the energy
eigenstates. In particular it allows to control unwanted
truncation effects.

For small $L$ one is confined to the first plateau only.
Translating a $L$-independent constant energy $W=m_\eta$ into a phase
shift we get a variety of $\delta$-values, all for a constant
$k = \ot \sqrt{m_\eta^2 - 4 m_\varphi^2} $, just what we expect
for a zero width resonance: a stable particle state.
For a naive estimator of the resonance energy one therefore might
consider just the single particle operator,
ignoring the states of higher energy coming from
above. However, increasing the lattice size would not improve the
result (In fact, the resonance is no asymptotic state of the theory).
There can be no avoided level crossing, as there are no 2-particle
states. Thus there will be no information on the resonance width.
On the other hand, ignoring the $\eta$-state and considering {\em only}
$\varphi \varphi$ states, is insufficient, too.
It is of crucial importance to
include all operators (with the quantum numbers of that sector) with
energies close to the considered energy eigenstates.
For further discussions, in particular with regard to the situation in
QCD see \cite{Lu91b}.

How can we understand the limit $g\to 0$? The energy gap becomes
smaller and the level crossing region narrower. The phase shift
approaches step function behaviour. In the Ising limit the
phase shift is then compatible with a constant (mod $\pi$).

\vspace{0.5cm}
{\bf Acknowledgment:}
We have benefitted much from discussions with H. Gausterer, M.
L\"uscher, M. Salmhofer and E. Seiler, and we want to thank.
\newpage


\newpage

{\noindent \Large \bf Tables}

\vspace{0.5cm}
{\noindent \bf Table 1:~}
This table summarizes the parameters for our simulations
together with the measured values of masses and resonance width
(in units of the lattice constant). We also show wave
function renormalization constants and the effective 3-point couplings.

\smallskip
\begin{center}
\begin{tabular}{|l|l|l|l|}
\hline
Model                 &I             &II            &III \\
\hline
$g$                   &0.            &0.02            &0.04 \\
$\kappa_\varphi$      &0.3948        &0.3897          &0.3700 \\
$\kappa_\eta$         &0.3268        &0.3323          &0.3700 \\
\hline
$m_\varphi$           &0.189921      &0.18965(13)     &0.19014(13) \\
$m_\eta$              &0.499545      &0.5008(4)       &0.5112(3) \\
$\Gamma_\eta$         &0.            &0.0047(4)       &0.0100(3) \\
$Z_\varphi$           &0.722(2)      &0.706(4)        &0.671(4) \\
$Z_\eta$              &0.849(3)      &0.816(5)        &0.619(4) \\
$\geff$     	      &0.            &0.0568(6)       &0.0938(8) \\
\hline
\end{tabular}
\end{center}

\newpage
\section*{Figures}

\vspace{0.5cm}
{\noindent \bf Fig. 1:~}
The dispersion relation of our single particle propagator (circles,
model (I) for $L=50$) approaches the continuum behaviour (broken curve)
for small $p_{1,\nu}$; it follows nicely the lattice propagator of a
free Gaussian particle (full curve).

\vspace{0.5cm}
{\noindent \bf Fig. 2:~}
We plot our values for $ \log [(M ( L ) - m) \sqrt{L}]$ vs. $L$, model
(I); these results for the single particle mass follow the expected
exponential dependence on the spatial size (\ref{mdecrease}).

\vspace{0.5cm}
{\noindent \bf Fig. 3:~}
(a) The 2-particle energy levels for a free Gaussian model
($\delta = 0$) as functions of the lattice size $L$. The dotted line
indicates the energy of a stable $\eta$ with $m_\eta =0.5$.
(b) The measured energy levels for the $\varphi\varphi$ and $\eta$
channel for the coupling constant values $g=0$ vs. the spatial lattice
extension $L$. The full curve shows the theoretical expectations for
the values from (\ref{Phaseshiftrelation}) for $\dising=-\frac{\pi}{2}$,
the horizontal line denotes the $\eta$ mass level. The dashed lines
indicate the 2- and the 4-particle thresholds.

\vspace{0.5cm}
{\noindent \bf Fig. 4:~}
The results of the phase shift for model (I), i.e. corresponding
to the energy data in fig. 3b, vs. $k/m_\varphi$. The full curve gives
the theoretical expectation $\dising=-\frac{\pi}{2}$. The dashed line
indicates the 4-particle threshold.

\vspace{0.5cm}
{\noindent \bf Fig. 5:~}
The measured energy levels for the $\varphi\varphi$ channel for the
coupling constant values (a) $g=0.02$ and (b) $g=0.04$, vs. the spatial
lattice extension $L$. The full curves show the theoretical expectations
for the values from (\ref{Phaseshiftrelation}) and $\dtotal$ as
discussed in the text, eq.(\ref{W}). The dashed lines indicate the 2-
and the 4-particle thresholds.

\vspace{0.5cm}
{\noindent \bf Fig. 6:~}
The phase shift as determined from the energy levels in figs. 5a and
5b vs. $k/m_\varphi$; the full curves denote the theoretical
expectations as discussed in the text. The dashed line indicates the
4-particle threshold.

\vspace{0.5cm}
{\noindent \bf Fig. 7:~} Contributions of the operators $N_1 , N_2$
and $N_3$ (circles, triangles, squares) to the energy eigenstates
(a) $|1\rangle$ , (b) $|2\rangle$ and  (c) $|3\rangle$, for case
(III), displayed as a function of the lattice size $L$.

\vspace{0.5cm}
{\noindent \bf Fig. 8:~} Contributions $\Big(\zeta_n^{(6)}\Big)^2$ of
the considered operators $N_n$ to the energy eigenstate $|6\rangle$ at
$L=60$ for model (II).

\end{document}